\documentclass[12pt]{article}
\usepackage{amsfonts}
\usepackage{amssymb}
\usepackage{amscd}
\def\D{\hbox{D\kern-.73em\raise.25ex\hbox{-}\raise-.25ex\hbox{ }}}
 \def\d{\hbox{d\kern-.33em\raise.75ex\hbox{-}\raise-.75ex\hbox{}}}

\def\GGG{\frak G }
\def\gr3{\GGG\,(\SSS_3)}

\def\gr2{\GGG\,(\SSS_2)}

\def\SSS{\frak S}

\def\al{{\alpha}}
\def\bet{{\beta}}

\def\gam{{\gamma}}

\def\ve{\varepsilon }

\def\vp{\vspace}
\def\hp{\hspace}
\def\ed{\end{document}}
\def\beq{\begin{equation}}
\def\eeq{\end{equation}}
\def\bea{\begin{eqnarray}}
\def\eea{\end{eqnarray}}
\def\ba{\begin{array}}
\def\ea{\end{array}}
\def\bi{\begin{itemize}}
\def\ei{\end{itemize}}

\def\lb{\label}

\def\nn{\nonumber}

\newcommand{\bp}{\noindent\begin{minipage}[c]}
\newcommand{\ep}{\end{minipage}}

\begin{document}


\title{ Path  Integral Approach to \\ Noncommutative Quantum Mechanics}

\author{ Branko Dragovich  \\
{\it Institute of Physics,  P.O.Box 57, 11001 Belgrade }\\ {\it
Serbia and Montenegro } \\ E-mail: dragovich@phy.bg.ac.yu \\
\\ {Zoran Raki\'c}\\
{\it Faculty of Mathematics, University of Belgrade} \\
{\it Studentski trg 16, P.O.Box 550, 11001 Belgrade} \\ {\it
Serbia and
Montenegro} \\
E-mail: zrakic@matf.bg.ac.yu}

\date{}

\maketitle

\begin{abstract}{ We consider Feynman's path integral approach to
quantum mechanics with a noncommutativity in position and momentum
sectors of the phase space. We show that a quantum-mechanical
system with this kind of noncommutativity  is equivalent to the
another one with usual commutative coordinates and momenta. We
found connection between quadratic classical Hamiltonians, as well
as Lagrangians, in their commutative and noncommutative regimes.
The general procedure to compute Feynman's path integral on this
noncommutative phase space with quadratic Lagrangians
(Hamiltonians) is presented. Using this approach, a particle in a
constant field, ordinary and inverted harmonic oscillators are
elaborated in detail.}
\end{abstract}

\section{ INTRODUCTION }

Quantum theories with noncommuting position and momentum
coordinates have been investigated intensively during the recent
years. Most of the research has been devoted to noncommutative
(NC) field theory (for a review, see e.g. \cite{nekrasov}).
Noncommutative quantum mechanics (NCQM) has been also
investigated, because it can be regarded as the corresponding
one-particle nonrelativistic sector of NC quantum field theory and
with relevance to concrete quantum systems.

 In a very general NCQM  one
 has that not only  $ [ \hat{x_a},\hat{p_b}] \neq 0$,
 but also $ [\hat{x_a},\hat{x_b}] \neq 0$ and
$[\hat{p_a},\hat{p_b}] \neq 0 . $ We consider here $D$-dimensional
NCQM which is based on the following algebra: \begin{equation} [
\hat{x_a},\hat{p_b}] = i\,\hbar\, ( \delta_{ab} - \frac{1}{4}\,
\theta_{ac}\, \sigma_{cb}), \quad [\hat{x_a},\hat{x_b}] =
i\,\hbar\, \theta_{ab},\quad [\hat{p_a},\hat{p_b}] =i\,\hbar\,
\sigma_{ab} , \label{1.1} \end{equation} where
${\Theta}=(\theta_{ab})$ and ${\Sigma} =(\sigma_{ab})$ are the
antisymmetric matrices with constant elements.

We investigate the Feynman path integral \cite{feynman} approach
to NCQM,
\begin{equation} {\mathcal K} (x'',t'';x',t')
=\int_{x'}^{x''} \exp \left ( \frac{i}{\hbar}\, \int_{t'}^{t''}
L(\dot{q},q, t)\, dt \right )\, {\mathcal D}q \,, \label{1.2}
\end{equation}  where ${\mathcal K}(x'',t'';x',t')$ is the kernel of the
unitary evolution operator $U (t)$ and $x''=q(t''), \ x'=q(t')$
are end points.
 In ordinary quantum mechanics (OQM) Feynman path integral for quadratic Lagrangians  can be
evaluated analytically and the exact form   \cite{steiner} is
\begin{equation}  {\mathcal K}(x'',t'';x',t') =\frac{1}{(i h)^{\frac{D}{2}}}
\sqrt{\det{\left(-\frac{\partial^2 {\bar S}}{\partial x''_a
\partial x'_b} \right)}} \exp \left(\frac{2\pi i}{h}\,{\bar
S}(x'',t'';x',t')\right), \label{1.3} \end{equation} where $ {\bar
S}(x'',t'';x',t')$ is the action for the classical trajectory.

 In this article we search the form of an effective quadratic  Lagrangian which
corresponds to a system with noncommutative phase space
coordinates (\ref{1.1}). This is necessary to know before to
employ Feynman's path integral method in NCQM. To this end, let us
note that algebra (\ref{1.1}) of operators $\hat{x_a},\,
\hat{p_b}$ can be replaced by the equivalent one \begin{equation}
[ \hat{q_a},\hat{k_b}] = i\,\hbar\, \delta_{ab}, \quad
[\hat{q_a},\hat{q_b}] = 0,\quad [\hat{k_a},\hat{k_b}] =0\, ,
\label{1.4} \end{equation} where linear transformations
\begin{equation} \hat{ x_a} = \hat{q_a} - \frac{\theta_{ab}\,
\hat{k_b}}{2}, \quad \hat{ p_a} = \hat{k_a} + \frac{\sigma_{ab}\,
\hat{q_b}}{2} \label{1.5} \end{equation} are used and summation
over repeated indices is assumed. According to (\ref{1.1}),
(\ref{1.4}) and (\ref{1.5}), NCQM related to the quantum phase
space $(\hat{p} ,\, \hat{x})$ can be regarded as an OQM on the
standard phase space $(\hat{k} ,\, \hat{q})$.  Similar aspects of
noncommutativity with $\sigma_{ab} =0$ are considered in
\cite{dragovich1} and \cite{dragovich2} .

\section{ DYNAMICS ON NONCOMMUTATIVE PHASE SPACE AND PATH INTEGRAL}

Let the most general  quadratic Lagrangian for a $D$-dimensional
system be
\begin{equation}  L(\dot{x}, x,t) = \frac{1}{2}\,\left(
\dot{x}^T \, \alpha\, \dot{x} + \dot{x}^T \, \beta\, x  + x^T \,
\beta^T\, \dot{x} + x^T \, \gamma\, x \right)+ \delta^T\, \dot{x}
+ \eta^T\, x + \phi , \label{2.1} \end{equation} where
coefficients of the $D\times D$ matrices $\alpha
=((1+\delta_{ab})\, \alpha_{ab}(t)),\ \beta =(\beta_{ab}(t)),\
\gamma =((1+\delta_{ab} )\, \gamma_{ab}(t)),$ $D$-dimensional
vectors $ \delta =(\delta_{a}(t)),$ \ $\eta =(\eta_{a}(t))$ and a
scalar $\phi =\phi(t)$ are some analytic functions of  the time
$t$. Matrices $\alpha$ and $\gamma$ are symmetric,  $\alpha$ is
nonsingular $(\det\alpha \neq 0)$ and index ${}^T$ denotes
transpose map.

The  Lagrangian (\ref{2.1}) can be presented in the  more compact
form: \begin{equation}    L(X,t) = \frac{1}{2}\, X^T \, M\, X +
N^T \,  X   + \phi , \label{2.2}
\end{equation} where $2D\times 2D$ matrix $M$ and $2D$-dimensional vectors
$X,\, N$ are defined as \bea & {M } = \left(
\begin{array}{ccc}
 \alpha & \beta  \\
 \beta^T & \gamma \end{array}
\right)\, , \quad  X^T = (\dot{x}^T\, ,\,  x^T) \, , \quad N^T =
(\delta^T\, ,\,  \eta^T) . \label{2.3} \eea Using the equations $
p_a = {\partial L \over \partial \dot x_a},$ one can express
${\dot x }$ as $ {\dot x } =  {\alpha^{-1}}\, ({p}  - {\beta}\, {x
} - \delta ). $ Since $\dot{x}$ is linear in $p$ and $x$, the
corresponding classical Hamiltonian $ H(p,x,t)= {p}^T\, \dot {x }
- L(\dot{x},x,t)$ becomes also quadratic, i.e. \beq \label{2.4}
H(p,x,t) = \frac{1}{2}\,\left( {p}^T \, A\, {p} + p^T \, B\, x  +
x^T \, B^T\, p  +  x^T \, C\, x \right) + D^T\, p + E^T\, x  + F ,
\end{equation}
where: \bea
  & & {A } =  {\alpha}^{-1}, \hspace{1.2cm}   {B }
=-\, {\alpha}^{-1}\, {\beta},  \hspace{1.2cm}  {C} =
{\beta}^T\, {\alpha}^{-1}\, {\beta} - {\gamma } , \nn \\
& &  {D} =- \, {\alpha }^{-1}\,  \delta, \hspace{0.5cm}  {E } =
\beta^T \, \alpha^{-1}\,  \delta - \eta , \hspace{0.5 cm} {F} =
{1\over 2}\, {\delta}^T \, {\alpha}^{-1} {\delta}\,  - {\phi }
\label{2.5} \,. \eea

Due to the symmetry of matrices $\alpha$ and $\gamma$ one has that
matrices $A = ((1+\delta_{ab})\, A_{ab}(t))$ and $C =
((1+\delta_{ab})\, C_{ab}(t))$ are also symmetric ($A^T = A ,\, \,
C^T = C$). The nonsingular $(\det {\alpha}\neq 0)$ Lagrangian
$L(\dot{x},x,t)\, $   implies nonsingular $ (\det{ A}\neq 0) $
Hamiltonian $H(p,x,t) $. Note that the inverse map, i.e. $H \to
L$, is given by the same relations (\ref{2.5}).

The Hamiltonian (\ref{2.4}) can be also presented in the compact
form \bea \hspace{-5mm} H(\Pi,t) = \frac{1}{2}\, \Pi^T \,
{\mathcal M}\, \Pi + {\mathcal N}^T \,  \Pi   + F , \label{2.6}
\eea where matrix ${\mathcal M}$ and  vectors $\Pi,\, {\mathcal
N}$ are \bea & {\mathcal M } = \left(
\begin{array}{ccc}
 A & B  \\
 B^T & C \end{array}
\right)\, , \quad  \Pi^T = (p^T\, ,\,  x^T) \, , \quad {\mathcal
N}^T = (D^T\, ,\,  E^T) . \label{2.7} \eea

One can easily show that \beq {\mathcal M}= \sum_{i =1}^3
\Upsilon_i^T(M)\,M\, \Upsilon_i(M) , \eeq where  \bea \nn
\Upsilon_1(M)=\left( \ba{lr} \al^{-1}  &0\\0 & -I\ea\right),  \,
\Upsilon_2(M)=\left( \ba{lr} 0\  &  \alpha^{-1} \beta\\
0 & 0\ea\right), \,  \Upsilon_3(M)=\left( \ba{lr} 0\  &0\\0 & \
 i\sqrt{2}\,I\ea\right) .\eea One has also  ${\mathcal N}= Y(M) \, N,  $
 where \bea\label{}  Y (M)= \left(\ba{rr}
-\,\al^{-1} & 0  \\ \bet^T\,\al^{-1}& -I  \ea\right) = -\Upsilon_1
(M) + \Upsilon^T_2 (M) + i \sqrt{2} \Upsilon_3 (M)\, \eea and $F =
N^T\, Z(M)\, N - \phi ,$ where \bea  Z(M ) = \left(\ba{ll}
\frac{1}{2}\, \al^{-1} & 0  \\ 0 & 0  \ea\right) =\frac{1}{2}\,
\Upsilon_1 (M) - \frac{i}{2\sqrt{2}}\, \Upsilon_3 (M)\, . \eea

 Eqs. (\ref{1.5}) can be rewritten in the compact form as \bea
\hat{\Pi} = \Xi \,\, \hat{K} , \quad \Xi = \left(
\begin{array}{ccc}
 I & \frac12\,\, {\Sigma}  \\
- \frac12\,\, {\Theta} & I \end{array} \right)\, , \quad \hat{K}=
\left(
\begin{array}{ccc}
 \hat{k}  \\
 \hat{q} \end{array}
\right) .\  \label{2.8} \eea

Since Hamiltonians depend on canonical variables, the
transformation  (\ref{2.8}) lead to the transformation of
Hamiltonians (\ref{2.4}) and (\ref{2.6}). To this end, let us
quantize the Hamiltonian (\ref{2.4}) and it easily becomes $
H(\hat{p},\hat{x},t) =  \frac{1}{2}\, ( {\hat{p}}^T \, A\,
\hat{p}\break + \hat{p}^T \, B\, \hat{x} + \hat{x}^T \, B^T\,
\hat{p} + \hat{x}^T \, C\, \hat{x} ) + D^T\, \hat{p} + E^T\,
\hat{x} + F  $ because (\ref{2.4}) is already written in the Weyl
symmetric form.

Performing linear transformations (\ref{1.5}) in the above
Hamiltonian we again obtain quadratic quantum Hamiltonian \bea
H_{\theta\sigma}(\hat{k},\hat{q},t) &=& \frac{1}{2}\,\left(
{\hat{k}}^T \, A_{\theta\sigma}\, \hat{k} + \hat{k}^T \,
B_{\theta\sigma}\, \hat{q} + \hat{q}^T \, B^T_{\theta\sigma}\,
\hat{k} + \hat{q}^T \, C_{\theta\sigma}\, \hat{q} \right) \nn \\
&& + D^T_{\theta\sigma}\, \hat{k} + E^T_{\theta\sigma}\, \hat{q} +
F_{\theta\sigma}\, ,\label{2.9} \eea where    \beq \ba{ll}
\displaystyle{{A}_{\theta\sigma} = {A} - {1\over 2}\,\, {B}\, {
\Theta} + {1\over 2}\,\, {\Theta}\, {B}^T -{1\over 4}\,\,
{\Theta}\, {C}\, { \Theta} ,} & \hp{10mm}
\displaystyle{{D}_{\theta\sigma} = {D} + {1\over
2}\,\, {\Theta}\, {E}}, \vp{2mm} \\
\displaystyle{{B}_{\theta\sigma} = {B} + \frac{1}{2}\, {\Theta}\,
{C} + \frac{1}{2}\, A\, \Sigma + \frac14\,\, \Theta\, B^T \,
\Sigma} , & \hp{10mm} \displaystyle{{E}_{\theta\sigma} = {E}-
\frac 12\, \Sigma\, D} ,  \vp{2mm}\\
\displaystyle{{C}_{\theta\sigma} = {C} - {1\over 2}\,\Sigma\, {B}
+ {1\over 2}\, B^T \, {\Sigma} -{1\over 4}\, {\Sigma}\, {A}\,
\Sigma}, & \hp{10mm} \displaystyle{{F}_{\theta\sigma} = {F}}
.\ea\label{2.10} \eeq  Note that for the nonsingular Hamiltonian $
H(\hat{p},\hat{x},t)$ and for sufficiently small $\theta_{ab}$ the
Hamiltonian $ H_{\theta\sigma}(\hat{k},\hat{q},t)$ is also
nonsingular. Classical analogue of (\ref{2.9}) maintains the same
form $ H_{\theta\sigma}(k,q,t) = \frac{1}{2}\, ( k^T \,
A_{\theta\sigma}\, {k} + {k}^T \, B_{\theta\sigma}\, {q} + {q}^T
\, B^T_{\theta\sigma}\, {k} \break  + {q}^T \, C_{\theta\sigma}\,
{q} )+ D^T_{\theta\sigma}\, {k} + E^T_{\theta\sigma}\, {q} +
F_{\theta\sigma}\,.$

In the more compact form, Hamiltonian (\ref{2.9}) is \bea  \hat
H_{\theta\sigma}(\hat{K},t) = \frac{1}{2}\,\, \hat K^T \,
{\mathcal M}_{\theta\sigma}\, \hat K + {\mathcal
N}_{\theta\sigma}^T \, \hat K + F_{\theta\sigma} , \label{2.11}
\eea where $2D\times 2D$ matrix ${\mathcal M}_{\theta\sigma}$ and
$2D$-dimensional vectors $\hat{K},\, {\mathcal N}_{\theta\sigma}$
are \bea & {\mathcal M}_{\theta\sigma} = \left(
\begin{array}{ccc}
 A_{\theta\sigma} & B_{\theta\sigma}  \\
 B_{\theta\sigma}^T & C_{\theta\sigma} \end{array}
\right)\, , \quad  \hat{K}^T = (\hat{k}^T\,, \,  \hat{q}^T) \, ,
\quad {\mathcal N}_{\theta\sigma}^T = (D_{\theta\sigma}^T\,, \,
E_{\theta\sigma}^T) . \label{2.12} \eea  From (\ref{2.6}),
(\ref{2.8}) and (\ref{2.11})  one can find connections between
${\mathcal M}_{\theta\sigma},\, {\mathcal N}_{\theta\sigma} , \, {
F}_{\theta\sigma} $ and ${\mathcal M},\, {\mathcal N}, \, F$,
which are given by the following relations: \bea\label{2.13} &&
\hp{0mm} {\mathcal M}_{\theta\sigma}=\Xi^T\,{\mathcal M}\,\, \Xi ,
\qquad  {\mathcal N}_{\theta\sigma}=\Xi^T\,{\mathcal N} , \qquad
{F}_{\theta\sigma} = F. \eea \vp{1mm}

To compute a path integral one can start from  its Hamiltonian
formulation on the phase space. However, such path integral on a
phase space can be reduced to the Lagrangian path integral on
configuration space whenever Hamiltonian is a quadratic polynomial
with respect to momentum $p$ (see, e.g. \cite{dragovich2}). Hence,
we need the corresponding classical Lagrangians  related to the
Hamiltonians (\ref{2.9}) and (\ref{2.11}). Using equations $
\dot{q}_a = \frac{\partial H_{\theta\sigma}}{\partial p_a} $ which
give $ k = A_{\theta\sigma}^{-1}\, (\dot{q} - B_{\theta\sigma}\, q
- D_{\theta\sigma}) $ we can pass from Hamiltonian (\ref{2.9}) to
the corresponding Lagrangian by relation $ L_{\theta\sigma}
(\dot{q},q,t) = k^T \dot{q}  - H_{\theta\sigma} (k,q,t) . $ Note
that coordinates $q_a$ and $x_a$ coincide when $\theta =\sigma
=0$. Performing necessary computations we obtain
 \bea  L_{\theta\sigma}(\dot{q}, q,t) &=&
\frac{1}{2}\,\left( \dot{q}^T \, \alpha_{\theta\sigma}\, \dot{q} +
\dot{q}^T \, \beta_{\theta\sigma}\, q  +  q^T \,
\beta^T_{\theta\sigma}\, \dot{q}  +  q^T \,
\gamma_{\theta\sigma}\, q \right) \nn \\ && +
\delta^T_{\theta\sigma}\, \dot{q} + \eta^T_{\theta\sigma}\, q +
\phi_{\theta\sigma} , \label{2.14} \eea or in the  compact form:
\bea   \hspace{5mm} L_{\theta\sigma}(Q,t) = \frac{1}{2}\, Q^T \,
M_{\theta\sigma}\, Q+ N_{\theta\sigma}^T \, Q +
\phi_{\theta\sigma} , \label{2.15} \eea where  \bea &
{M_{\theta\sigma}}  = \left(
\begin{array}{ccc}
 \alpha_{\theta\sigma} & \beta_{\theta\sigma}  \\
 \beta_{\theta\sigma}^T & \gamma_{\theta\sigma} \end{array}
\right)\, , \quad   Q^T = (\dot{q}^T\,, \,  q^T) \, , \quad N^T =
(\delta_{\theta\sigma}^T\,, \,  \eta_{\theta\sigma}^T).
\label{2.16} \eea Then the connection between ${M_{\theta\sigma}},
{N_{\theta\sigma}}, \phi_{\theta\sigma} \, $ and $\, M,N,\phi$ are
given by the following relations: \beq \label{2.17} \hp{-15mm}
\ba{l} M_{\theta\sigma} = \sum\limits_{i,j=1}^3 \, \Xi_{ij}^T\, \,
M\, \,\Xi_{ij},\quad \quad
  \Xi_{ij}=\Upsilon_i(M)\,\,\Xi\,\,\Upsilon_j(\mathcal{M}_{\theta\sigma})
, \ea \eeq  and \beq \label{2.18}  N_{\theta \sigma} = Y
(\mathcal{M}_{\theta\sigma})\, \Xi^T\, Y(M)\, N, \quad
\phi_{\theta \sigma} = \mathcal{N}_{\theta\sigma}^T \, Z
(\mathcal{M}_{\theta\sigma}) \, \mathcal{N}_{\theta\sigma} - F \,
. \eeq

 In more detail, the connection between  coefficients of the
 Lagrangians $L_{\theta\sigma}$
 and  $L$ is given by the relations:
 \bea\nn   {\al}_{\theta\sigma} & =& \big[\,
{\al}^{-\,1} - \frac 12 \,\, (\Theta\, {\bet}^{T}\, {\al}^{-\,1}-
{\al}^{-\,1}\, {\bet}\, \Theta) -\frac14\,\,\Theta\,( {\bet}^{T}\,
{\al }^{-\,1}\, {\bet} - \gam)
\,\Theta \, \big]^{-\,1}\,, \nn  \\
 \nn   {\bet}_{\theta\sigma} & = & {\al}_{\theta\sigma}\,
 \big( {\al}^{-\,1}\,
 {\bet}  - \frac 12 \,\,( {\al}^{-\,1}\,\Sigma -\Theta\, {\gam}+
  \Theta\, {\bet}^{T}\, {\al}^{-\,1}\,{\bet})
  +\frac14\,\,\Theta\,{\bet}^{T}\,{\al}^{-\,1}\,\Sigma\big)\,, \vp{2mm} \\
\nn   {\gam}_{\theta\sigma} &=& \gamma +  \bet_{\theta\sigma}^T \,
\al_{\theta\sigma}^{-\,1} \, \bet_{\theta\sigma} -  {\bet }^{T}\,
{\al}^{-\,1}\, {\bet}  + \frac 14\,\, \Sigma\,{\al}^{-\,1}\,
\Sigma\, \\\nn  & &  -\frac 12\,\,(\Sigma\,{\al}^{-\,1}\,\bet
-{\bet }^{T}\,
{\al}^{-\,1}\,\Sigma ) \,, \\
 {\delta}_{\theta\sigma} & = &
{\al}_{\theta\sigma}\,\big({\al}^{-\,1}\,{\delta}+\frac
12\,\,(\Theta\,\eta- \Theta\,{\bet}^{T}\,
{\al}^{-\,1}\,{\delta}) \big)\,, \nn  \\
 {\eta }_{\theta\sigma}& = & \eta + \bet_{\theta\sigma}^T \,
\al_{\theta\sigma}^{-\,1} \, \delta_{\theta\sigma} -
{\bet}^{T}\,{\al}^{-\,1}\, {\delta}  - \frac 12\,\,
\Sigma\,{\al}^{-\,1}\,{\delta}\,,\nn  \\
 {\phi }_{\theta\sigma} & = & \phi + \frac 12\,\,
{\delta}_{\theta\sigma}^T\,  \al_{\theta\sigma}^{-\,1} \,
\delta_{\theta\sigma} - \frac 12\,\, \delta^T\,
{\al}^{-\,1}\,{\delta} \,.
 \label{2.17}\eea
Note that $\alpha_{\theta \sigma}, \,  \delta_{\theta \sigma} $
and $ \phi_{\theta \sigma}$ do not depend on $\sigma$.

If we know Lagrangian (\ref{2.1}) and algebra (\ref{1.1}) we can
obtain the corresponding effective Lagrangian (\ref{2.14})
suitable for the path integral in NCQM. Exploiting the
Euler-Lagrange equations $  \frac{\partial
L_{\theta\sigma}}{\partial q_a} -\frac{d}{dt}  \frac{\partial
L_{\theta\sigma}} {\partial{\dot q}_a} =0 $ one can obtain
classical trajectory $q_a =q_a (t)$ connecting  end points $x' =
q(t')$ and $x''= q(t'')$, and the corresponding  action ${\bar
S}_{\theta\sigma} (x'',t'';x',t') =\int_{t'}^{t''}
L_{\theta\sigma} (\dot{q}, q,t)\, dt$. Path integral in NCQM is a
direct analogue of (\ref{1.3}) and its exact expression in the
form of quadratic actions ${\bar S_{\theta\sigma}}(x'',t'';x',t')$
is \bea  {\mathcal K}_{\theta\sigma}(x'',t'';x',t') = \frac{1}{(i
h)^{\frac{D}{2}}} \sqrt{\det{\left(-\frac{\partial^2 {\bar
S_{\theta\sigma}}}{\partial x''_a\,
\partial x'_b} \right)}}  \exp \left(\frac{2\,\pi\, i}{h}\,{\bar
S_{\theta\sigma}}(x'',t'';x',t')\right). \label{e3.1}  \eea

 \subsection{A particle in a constant field and on noncommutative phase space}

 The starting  Lagrangian is
 \bea\label{3.1} L(\dot{x},x) = \frac{m}{2}\, (\dot{x}_1^2 +
\dot{x}_2^2 ) -\eta_1\, x_1  - \eta_2\, x_2 . \eea

 Using  (\ref{2.17}), one can
easily find the Lagrangian $L_{\theta\sigma}(\dot{q},q,t)$ :
\bea\lb{e17c} \hp{-15mm} L_{\theta\sigma}  &=& \frac{m}{2} \, \, (
{\dot{q}_1}^2 +{\dot{q}_2}^2 ) +\frac{\sigma}{2} \,( q_1\,
\dot{q}_2 -  q_2\, \dot{q}_1 ) + \frac{m\, \theta}{2} \,( \eta_1\,
\dot{q}_2 + \eta_2\, \dot{q}_1 ) \nn \\ \hp{-15mm} && - \
\Big(1-\frac{\theta\,\sigma}{4}\Big) \big(\eta_1\, q_1  + \eta_2\,
q_2\big) +\frac{m\,\theta^2}{8}\,({\eta_1}^2+{\eta_2}^2)\,.
\label{3.2} \eea The corresponding equations of motion are $
\ddot{q}_1 -\xi\, \dot{q}_2 + \eta_1 \,\phi= 0, \, \, \ddot{q}_2 +
\xi\, \dot{q}_1 + {\eta_2}\,\phi = 0, $ where $
\xi=\frac{\sigma}{m}, \, \, \phi=\frac {1}{m}\,\Big( 1-
\frac{\theta\, \sigma}{4}\Big)\ .$ One can transform the above
system of equations to  $ {q}_1^{(3)}+ \xi^2\, \dot{q}_1+
\eta_2\,\xi\, \phi= 0, \quad {q}_2^{(3)}+ \xi^2\, \dot{q}_2 -
\eta_1\,\xi\, \phi = 0, $ Their solutions are:
 $ q_1(t)= C_1+ C_2\,\cos[\,\xi\,t\,] +C_3\,\sin[\,\xi\,t\,] -
\frac{\phi\,\eta_2}{\xi}\,t ,\quad q_2(t)=D_1+
D_2\,\cos[\,\xi\,t\,] +D_3\,\sin[\,\xi\,t\,] +
\frac{\phi\,\eta_1}{\xi}\,t ,$ where $C_1,C_2,C_3,D_1, D_2$ and
$D_3$ are constants. Since the functions $q_1$ and $q_2$ have to
satisfy coupled differential equations  one has $ D_2=C_3,\quad
D_3= -\, C_2. $ Employing the conditions
 $ q_1(0)=x_1',\quad q_1(T)=x_1'', \quad
q_2(0)=x_2',\quad  q_2(T)=x_2''$ one obtains: $\\ {} \quad \quad
\quad C_1= \frac{(x_1'+x_1'')\,\xi+\phi\,\eta_2\,T}{2\,\xi}-
\frac{\big((x_2'-x_2'')\,\xi+ \phi\,\eta_1\,T\big)
}{2\,\xi\,}\,\cot\Big[\,\frac{\xi\,T}{2}\,\Big]\,, \\ {} \quad
\quad \quad  C_3=
\frac{(x_2'-x_2'')\,\xi+\phi\,\eta_1\,T}{2\,\xi}-
\frac{\big((x_1'-x_1'')\,\xi-
\phi\,\eta_2\,T\big)}{2\,\xi}\,\cot\Big[\,\frac{\xi\,T}{2}\,\Big]\,,
\\ {} \quad \quad \quad  D_1= \frac{(x_2'+x_2'')\,\xi-\phi\,\eta_1\,T}{2\,\xi}-
\frac{\big((x_1''-x_1')\,\xi+
\phi\,\eta_2\,T\big)}{2\,\xi}\,\cot\Big[\,\frac{\xi\,T}{2}\,\Big]\,,
\\  {} \quad \quad \quad  D_3= \frac{(x_1''-x_1')\,\xi+\phi\,\eta_2\,T}{2\,\xi}-
\frac{\big((x_2'-x_2'')\,\xi+ \phi\,\eta_1\,T\big)}{2\,\xi}
\,\cot\Big[\,\frac{\xi\,T}{2}\,\Big]\,. $

The Lagrangian for classical trajectory is $ L_{\theta\sigma}(\dot
q,q) =  - \frac{m\, \,\xi\,\phi\,\,t}{2}\Big(
(C_3\,\eta_1+D_3\,\eta_2)\,\cos[\,\xi\,t\,] +
(D_3\,\eta_1-C_3\,\eta_2)\,\sin[\,\xi\,t\,]\Big)
 - \frac{m}{2}\Big(
(C_3\,D_1-C_1\,D_3)\, \xi^2+
(\theta\,\xi+3\,\phi)(C_3\,\eta_2-D_3\,\eta_1)\,\cos[\,\xi\,t\,] +
(C_1\,C_3+D_1\,D_3)\, \xi^2+
(\theta\,\xi+3\,\phi)(C_3\,\eta_1+D_3\,\eta_2)\,\sin[\,\xi\,t\,]\Big)
  -
\frac{m\,\phi}{2}\,\big(C_1\,\eta_1+D_1\,\eta_2\big) +
\frac{m\,(\theta\,\xi+2\,\phi)^2\,(\eta_1^2+\eta_2^2)}{8\,\xi^2}.
$ Using this Lagrangian we  compute the classical action \bea &&
{\bar S}_{\theta\sigma} (x'',T;x',0)= \int\limits_{0}^T \,
L_{\theta\sigma}(\dot{q},q)\, d\hspace{.1mm}t
 =
 \frac{m\,\xi}{2}\,\Big(x''_2\,x'_1-x''_1\,x'_2 \Big)+
 \frac{m\,\xi\cot\big[\,\frac{\xi\,T}{2}\,\big]}{4}\,
 \nn \\ && \nn
  \times \Big((x''_1-x'_1)^2 +(x''_2-x'_2)^2 \Big) +
 \frac{2\,\lambda+m\,\theta\,\xi-\lambda\,\xi\,T\cot\big[\,\frac{\xi\,T}{2}\,\big]}{2\,\xi}
 \nn \\ && \nn
  \times  \Big((x''_2-x'_2)\,\eta_1 -(x''_1-x'_1)\,\eta_2\Big) -
\frac{\lambda\,T}{2}\, \Big((x''_1+x'_1)\,\eta_1
+(x''_2+x'_2)\,\eta_2\Big)
  \\ &&
+\, \frac{T}{8\,m\,\xi^2}\,
\Big(2\,\xi\,\lambda^2\,T\cot\Big[\,\frac{\xi\,T}{2}\,\Big] +
\big(
m^2\,\theta^2\xi^2-4\,\lambda^2\big)\Big)\,(\eta_1^2+\eta_2^2)\, .
\label{3.3} \eea Finally, we have  \bea \hp{-10mm} & &  {\mathcal
K}_{\theta\sigma} (x'',T;x',0) =
 \frac{m\,|\hp{.2mm}\xi\,\big|}
 {2\,i\, h\,\big|\!\sin\big[\frac{\xi\,T}{2}\big]\hp{-.1mm}\big|}\,
 \exp\left(
 \frac{2\pi i}{h}\, \bar{S}_{\theta\sigma} (x'',T;x',0)   \right)
\label{3.4}.\eea

 \subsection{ Ordinary and inverted harmonic oscillator on noncommutative plane}

 The  Lagrangian in the question is \bea\lb{3.5}
 L(\dot{x},x) = \frac{m}{2}\, (\dot{x}_1^2  + \dot{x}_2^2) -\ve
\frac{m\,\omega^2}{2}\, (x_1^2 + x_2^2)\,, \eea where $\ve=1$ and
$\ve=-1$ are for ordinary and inverted harmonic oscillators,
respectively.  Using formulas (\ref{2.17}), one can easily find
the corresponding noncommutative Lagrangian \bea \nn
L_{\theta\sigma}( \dot{q},q) &=&
  \frac{m}{2\, \kappa} \,  \big(
{\dot{q}_1}^2 +{\dot{q}_2}^2\big)+ \frac{\sigma+\ve\,
m^2\,\omega^2\,\theta }{2\, \kappa} \, \big( \dot{q}_2\,q_1-
\dot{q}_1\, q_2\big) \\ &&
     - \ \frac{\ve\,m\,\omega^2}{2\, \kappa}\,
     \lambda^2 \big({q_1}^2 +{q_2}^2\big) \,, \label{3.6} \eea
where $\kappa = 1+\frac{\ve\,m^2\, \omega^2\, \theta^2}{4}$ and
$\lambda = 1-\frac{\theta\,\sigma}{4}$.

From (\ref{3.6}), we obtain the  Euler-Lagrange equations,
$\ddot{q}_1- \chi\, \dot{q}_2 + \mu\,q_1 =0 ,\quad \ddot{q}_2+
\chi\,\dot{q}_1+ \mu\,q_2 =0 ,$ where $\chi=
\frac{\sigma}{m}+\ve\, m\,\omega^2\, \theta\,, \quad
\mu=\ve\,\omega^2\,\lambda^2.$ Let us note that these equations
form a coupled system of second order differential equations,
which is more complicated than in the commutative case
($\theta=\sigma=0$). One can transform this system  to
${q}_1^{(4)}+ \big( \chi^2+2\,\mu \big)\,{q}_1^{(2)}+ \mu^2\,q_1
=0 ,\quad
 {q}_2^{(4)}+ \big( \chi^2+2\,\mu \big)\,{q}_2^{(2)}+
\mu^2\,q_2 =0. $ The solution of these equations  for
$\big|\theta\, \sigma\big| <4,\, \omega>0\,,\chi\neq 0$ and $\ve
> 0$ has the form $\left( \nu=\frac{\chi^2+4\,\mu}{4} \right) $ :
\beq\lb{3.7} \ba{l} \displaystyle{{q}_1(t) = C_1 \cos [\,
\frac{\chi}{2}\,t\,]\cos [ \sqrt{\ve\,|\nu|\,}\,t\,] + C_2
\cos [\, \frac{\chi}{2}\,t\,]\Big|\sqrt{\ve}\,\sin [ \sqrt{\ve\,|\nu|\,}\,t\,]}\Big| \vp{1mm} \\
\displaystyle{\hp{-2mm} +\  C_3 \sin [\, \frac{\chi}{2}\,t\,]\cos
[ \sqrt{\ve\,|\nu|\,}\,t\,] +C_4 \sin [\,
\frac{\chi}{2}\,t\,]\Big|\sqrt{\ve}\,\sin[\sqrt{\ve\,|\nu|\,}\,t\,]\Big|\,,}  \vp{2mm}\\
 \displaystyle{{q}_2(t) = D_1 \cos [\, \frac{\chi}{2}\,t\,]\cos [
\sqrt{\ve\,|\nu|\,}\,t\,] + D_2
\cos [\, \frac{\chi}{2}\,t\,]\Big|\sqrt{\ve}\,\sin [ \sqrt{\ve\,|\nu|\,}\,t\,]\Big|} \vp{1mm} \\
\displaystyle{\hp{-2mm} +\ D_3 \sin [\, \frac{\chi}{2}\,t\,]\cos [
\sqrt{\ve\,|\nu|\,}\,t\,] +D_4 \sin [\,
\frac{\chi}{2}\,t\,]\Big|\sqrt{\ve}\,\sin [
\sqrt{\ve\,|\nu|\,}\,t\,]\Big|\,.} \ea\eeq

 Let us note that in the case of inverted harmonic oscillator
$(\ve =-1)$  the trigonometric functions in
$\sqrt{\ve\,|\nu|\,}\,t$ from (\ref{3.7}) become the corresponding
hyperbolic functions in $\sqrt{\mid\nu\mid}\,t\,$.
 After imposing connections between $q_1$ and $q_2$ given by
coupled differential equations, we obtain the following relations
between constants $C$ and $D$: $ D_1=C_2,\ \ D_2=-C_1,\ \
D_3=C_4,\ \ D_4=-C_3\,.$ The unknown constants $C_1,C_2,C_3$ and
$C_4$ one can find from end conditions $ q_1(0)=x_1',\quad
q_1(T)=x_1'', \quad q_2(0)=x_2',\quad  q_2(T)=x_2''$ . Then one
obtains the solutions \bea  \nn  {q}_1 &=& \frac{1}{\sin [
\sqrt{\ve\,|\nu|\,}\,\,T\,]} \Big(\! \Big(x_1''\, \cos\!
\Big[\frac{\,\chi}{2}\,(t-T)\Big] + x_2''\,
\sin\!\Big[\frac{\,\chi}{2}\,(t-T)\Big]\Big)\sin [
\sqrt{\ve\,|\nu|\,}\,\,t\,]\\
\label{3.8} & & -\, \Big(x_1'\, \cos\! \Big[\frac{\chi\,t}{2}\Big]
+ x_2'\, \sin\! \Big[\frac{\chi\,t}{2}\Big]\Big) \sin \big[
\sqrt{\ve\,|\nu|\,}\,(t-T)\big]\Big),\lb{} \\
\nn {q}_2 & = & \frac{1}{\sin [ \sqrt{\ve\,|\nu|\,}\,\,T\,]}
\Big(\! \Big(x_2''\, \cos\! \Big[\frac{\,\chi}{2}\,(t-T)\Big] -
x_1''\, \sin\! \Big[\frac{\,\chi}{2}\,(t-T)\Big]\Big)\sin [
\sqrt{\ve\,|\nu|\,}\,\,t\,]
\\
\label{3.9} &&  +\ \Big(- x_2'\, \cos\!
\Big[\frac{\chi\,t}{2}\,\Big] + x_1'\, \sin\!
\Big[\frac{\chi\,t}{2}\,\Big]\Big) \sin \big[
\sqrt{\ve\,|\nu|\,}\,(t-T)\,\big]\Big) .  \eea Inserting the above
expressions and their time derivatives into (\ref{3.6}) we find
\bea && \nn L_{\theta\sigma}(\dot q,q) = \frac{m\,\ve\,
\nu}{2\,\kappa\ \sin^2\big[\sqrt{\ve\,|\nu|\,}\,T\,\big]} \bigg(
(x_1''^2+x_2''^2)\,\cos\big[\,2\sqrt{\ve\,|\nu|\,}\,t\,\big]+(x_1'^2+x_2'^2) \\
\hp{-24mm} \nn && \times\,
\cos\big[\,2\,\sqrt{\ve\,|\nu|\,}\,(t-T)\,\big] - 2\cos\big[
\sqrt{\ve\,|\nu|\,}\,(2\,t-T)\,\big]\Big(
(x_1'\,x_1''+x_2'\,x_2'')\\
\hp{-24mm}  && \times\,
\cos\Big[\,\frac{\chi\,T}{2}\,\Big]+(x_1''\,x_2'-x_1'\,x_2'')\,
\sin\Big[\,\frac{\chi\,T}{2}\,\Big]\Big)\bigg) . \label{3.9} \eea
Using (\ref{3.9}),  we finally compute the corresponding classical
action \bea && {\bar S}_{\theta\sigma}
(x'',T;x',0)=\int\limits_{0}^T \, L_{\theta\sigma}(\dot{q},q)\,
d\hspace{.1mm}t \label{3.10}
\\  \nn   &=& \frac{m\,\sqrt{\ve\,|\nu|\,}}{2\,\kappa\,
 \sin\big[\,\sqrt{\ve\,|\nu|\,}\,T\,\big]}
 \bigg(\!\!- 2\,\Big(
(x_1'\,x_1''+ x_2'\,x_2'') \cos \Big[\,\frac{\chi\,T}{2}\,\Big]
\\ \hp{-25mm} \nn  &+&
 (x_1'^2+x_1''^2+ x_2'^2+x_2''^2)\cos\big[\,\sqrt{\ve\,|\nu|\,}\,T\,\big]
 + 2\,(-x_1''\,x_2'+ x_1'\,x_2'') \sin \Big[\,\frac{\chi\,T}{2}\,\Big] \bigg)
 .  \eea If we take into account expression
(\ref{3.10}) then  we finally have  \bea
 \hp{-4mm}  {\mathcal K}_{\theta\sigma} (x'',T;x',0) =
 \frac{1}{i\, h} \frac{m\,\big|\sqrt{\ve\,|\nu|}\,\big|\,}
 {\kappa\,
 \big|\!\sin\big[\,\sqrt{\ve\,|\nu|}\,T\,\big]\hp{-.1mm}\big|}\,
 \exp\left(
 \frac{2\pi i}{h}\, \bar{S}_{\theta\sigma} (x'',T;x',0)   \right)
\label{3.11} .  \eea


\section{CONCLUDING REMARKS}

 Let us mention  that taking $\sigma=0 ,\, \,  \theta=0$ in
the above formulas we recover expressions for the Lagrangian
$L(\dot{q}, q)$, action $\bar{S} (x'',T;x',0) $ and probability
amplitude ${\mathcal K} (x'',T;x',0)$  of the ordinary commutative
case.

 Note that a similar path integral approach with $\sigma =0$
 has been considered
in the context of the Aharonov-Bohm effect , the Casimir effect ,
a quantum system in a rotating frame , and the Hall effect (for
references on these and some other related subjects, see
\cite{dragovich1,dragovich2}). Our investigation includes all
systems with quadratic Lagrangians (Hamiltonians).

On the basis of the expressions presented in this article, there
are many possibilities to discuss noncommutative
quantum-mechanical systems with respect to various values of
noncommutativity parameters $\theta$ and $\sigma$. For example, in
the case $\theta \sigma =4$ one has a critical point in the above
our particular two-dimensional models (see also
\cite{nersessian}).


\section*{Acknowledgments}

B.D. would like to thank the organizers of the {\it Fifth
International Workshop} "Lie Theory and its  Applications in
Physics" for invitation and hospitality, as well as for a
stimulating and pleasant atmosphere. A part of this paper was done
during stay of B.D. in the Steklov Mathematical Institute, Moscow.
The work on this article was partially supported by the Serbian
Ministry of Science, Technologies and Development under contracts
No 1426 and No 1646. The work of B.D. was also supported in part
by RFFI grant 02-01-01084.










\end{document}